\documentclass[journal]{IEEEtran}

\usepackage{array}
\usepackage{mdwmath}
\usepackage{mdwtab}
\usepackage{eqparbox}
\usepackage{url}
\usepackage{xcolor}
\usepackage{soul}
\usepackage{subfigure}
\usepackage{graphicx}
\usepackage{dcolumn}
\usepackage{bm}
\usepackage{float}
\usepackage{cite}
\usepackage{color}
\usepackage{amsmath}
\usepackage{resizegather}
\usepackage{multicol}
\usepackage{graphicx,epsfig,epstopdf}
\usepackage{lipsum}
\usepackage{mathtools}
\usepackage{cuted}

\begin{document}

\title{Analysis of Electromagnetic Scattering from Array of Time-Modulated Graphene Ribbons}

\author{Mahdi~Rahmanzadeh, Behzad Rejaei, and Amin Khavasi

\thanks{The authors are with the Department of Electrical Engineering, Sharif University of Technology, Tehran 11155-4363, Iran. Corresponding author is B. Rejaei (email:rejaei@sharif.edu)}}

\markboth{Journal of \LaTeX~ August~2022}%
{Shell \MakeLowercase{\textit{et al.}}: Bare Demo of IEEEtran.cls for IEEE Journals }

\maketitle

\begin{abstract} 
An accurate and fast method is presented for scattering of electromagnetic waves from an array of time-modulated graphene ribbons. We derive a time-domain integral equation for induced surface currents under subwavelength approximation. Using the method of harmonic balance, this equation is solved for a sinusoidal modulation. The solution of the integral equation is then used to obtain the transmission and reflection coefficients of time-modulated graphene ribbon array. The accuracy of the method was verified through comparison with results of full-wave simulations. In contrast with previously reported analysis techniques, our method is extremely fast and can analyze structures with a much higher modulation frequency. The proposed method also provides interesting physical insights useful for designing novel applications and opens up new vistas in the fast design of time-modulated graphene-based devices.
\end{abstract}


\IEEEpeerreviewmaketitle

\section{Introduction}

\IEEEPARstart{T}{emporal} modulation of electromagnetic structures provides an additional degree of freedom which may be employed to realize interesting wave effects such as frequency conversion, non-reciprocity, and signal amplification \cite{liu2018huygens, zhang2018space,caloz2019spacetime,hadad2015space, taravati2019generalized}. Based on these effects,  novel applications have been proposed in a wide range of frequencies from microwave to optical regions \cite{taravati2016mixer, pacheco2020temporal, taravati2022microwave, caloz2019spacetime2,sounas2017non,huidobro2019fresnel}. Dynamic modulation of a medium is usually achieved by means of acoustic waves \cite{yu2020acousto}, charge carrier generation using laser beams\cite{wen2014graphene}, varactor diodes \cite{liu2018huygens}, and mechanical effects\cite{ra2020nonreciprocal}. At terahertz frequencies, however,  a promising way of achieving time-modulated devices  is by using graphene-based structures.

Graphene, a two-dimensional (2D) layer of carbon atoms arranged in a honeycomb lattice, has already demonstrated unique mechanical, electric, magnetic, thermal, and optical properties, spurring tremendous interest \cite{geim2010rise,grigorenko2012graphene, novoselov2004electric}. These properties have made graphene an intriguing material for various applications in the last decade \cite{abdollahramezani2015analog, rahmanzadeh2018multilayer, momeni2018information,vakil2011transformation,chen2012terahertz}. Recently, time modulation and graphene were combined to achieve novel graphene-based applications \cite{menendez2017frequency,liu2016time, galiffi2019broadband, shirokova2019scattering, maslov2018temporal, menendez2019selective, correas2018magnetic,wang2020theory}. The surface carrier concentration of a graphene layer can be altered by an external gate voltage which changes the surface conductivity of graphene and has established graphene as a tunable material. Graphene surface conductivity can be dynamically modulated with frequencies up to several tens of gigahertz (GHz) \cite{li2014ultrafast,tasolamprou2019experimental}, which paves the way for various applications such as frequency comb\cite{menendez2017frequency}, isolators\cite{wang2020theory}, temporal wood anomalies \cite{galiffi2020wood}, secure communications\cite{sedeh2021active}, and circulators\cite{wang2020theory}.

Accurate, efficient, and fast analysis is critical to design and optimization of dynamically modulated graphene-based devices.  However, previously reported analysis methods for these structures have heavily relied on time-consuming full-wave simulations. Alternatively, adiabatic approximation is used in cases where the modulation frequency is much lower than the operating frequency. This, however,  restricts the usage of adiabatic approximation in many applications, particularly in harmonic frequency generation\cite{wang2020theory,menendez2019selective,salehi2022frequency,rajabalipanah2020reprogrammable,qing2021multifunctional}. A continuous time-varying (TV) graphene sheet can be analytically analyzed by expanding electromagnetic fields on both sides of graphene and then applying appropriate boundary conditions. Nevertheless, analysis of patterned graphene, which can show interesting phenomena not achievable in a continuous sheet (e.g., plasmonic resonance), is much more sophisticated. To the best of the authors’ knowledge, no rigorous and analytical or semi-analytical solution has yet been presented for patterned graphene with relatively fast temporal modulation.

In this paper, we study the scattering of a transverse-magnetic (TM) polarized electromagnetic wave by a periodic array of time-modulated graphene ribbons using a semi-analytical technique. We assume that at all frequencies involved (that of the incident wave and those additional frequencies generated by time modulation which have significant contribution), graphene ribbons are in a subwavelength regime. Then, we derive a semi-analytical expression for the surface current density induced on graphene ribbons under sinusoidal modulation. Next, the time-harmonic expansion of fields, in combination with the surface current density on the graphene ribbons, is used to determine the transmission and reflection coefficients of the structure. We validate the proposed method against full-wave numerical simulations. The results show that our proposed method is accurate and fast and can handle high modulation frequencies.  Physical explanations of electromagnetic response of the ribbon array are provided and discussed. Finally, the limitations of our proposed method are outlined.

The paper is organized as follows: in Sec. II, we present a semi-analytical method for calculation of the surface current density on dynamically modulated graphene ribbons in an infinite periodic array. Using the results obtained, reflection/transmission coefficients will next be computed. In section III, we present different numerical examples and validate the results obtained by comparison with results from full-wave electromagnetic simulations. Physical explanations are provided and limitations of the proposed method are discussed. The paper will be concluded in section IV.

\section{Scattering problem for a periodic array of time-modulated graphene ribbons}
\begin{figure}
\centering \includegraphics[width=\linewidth]{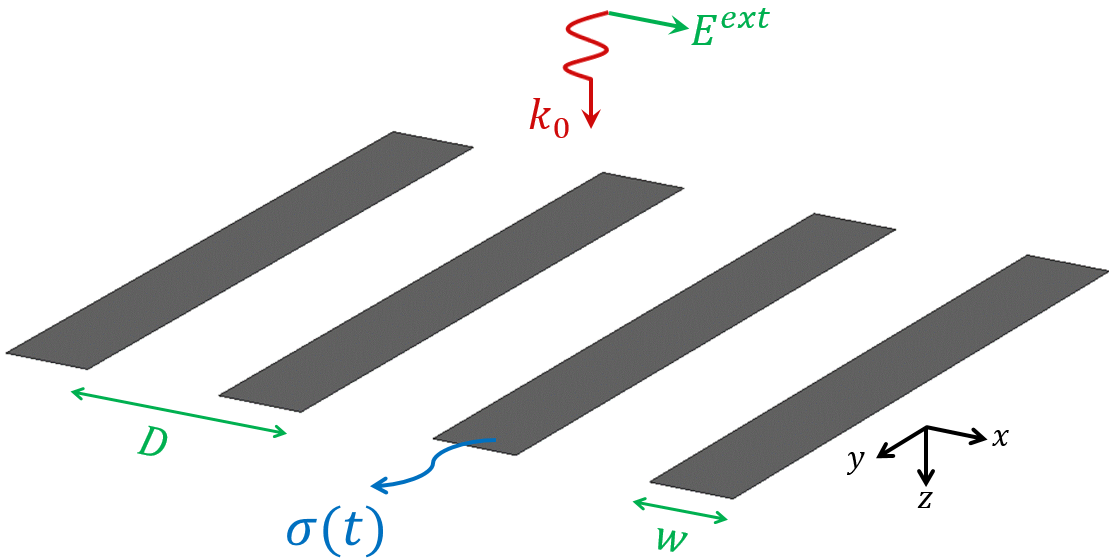}
\caption{Periodic array of time-modulated graphene ribbons. Graphene ribbons are illuminated by a monochromatic TM polarized wave.}
\label{GR_Fig}
\end{figure}

Fig.~\ref{GR_Fig} shows the structure under study: a periodic array of TV-graphene ribbons with lattice constant $D$. Graphene ribbons are suspended in free space, are infinitely long in the $y$-direction, and their width is $w$. The structure is illuminated by a normally incident monochromatic TM polarized plane wave (magnetic field along the $y$-direction) with the frequency $\omega$. Below, we shall first discuss the surface conductivity of a dynamically modulated graphene sheet. Next, equations will be derived for surface currents induced on TV graphene ribbons  by the incident wave. Finally, generalized reflection and transmission coefficients will be obtained for the ribbon array. A time dependence of the form exp$(j\omega t)$ is implicitly assumed throughout this paper.

\subsection{Surface conductivity of a TV-graphene sheet}

It is well-known that the surface conductivity of graphene can be tuned by applying a gate voltage. This enables the realization of time-modulated conductivity for the ribbons. We shall view graphene as a conductive sheet with a surface current $\mathbf{J}(t)$ that satisfies  the time-domain Drude equation \cite{galiffi2020wood}:
\begin{equation}
\frac{1}{W_D(t)}\left( \frac{\partial}{\partial t} +\gamma\right) \mathbf{J}(t) =\mathbf{E}(t)\label{Drude_TD}
\end{equation}
where
\begin{equation}
W_D(t)=\frac{E_F(t) e^2}{\pi\hbar^2}
\end{equation}
is the Drude weight quantifying the density of charge carriers at any given instant in time, $e=1.6\times10^{-19}$C is the electron charge, $E_F$ is the Fermi level energy, $\hbar$ is the reduced Planck’s constant, $\gamma$ is a phenomenological dissipation rate, and $\mathbf{E}(t)$ is the in-plane electric field. The loss rate can be described through $\gamma=1/\tau$= $e v^2_F/m E_F$, where $v_F=9.5\times10^7$cm/s is Fermi velocity of charge carriers and $m$ is electron mobility. The electron mobility of graphene on a substrate ranges from about $0.1 ~\textrm{m}^2/\textrm{Vs}$ to $6~ \textrm{m}^2\textrm{/Vs}$, depending on the fabrication process \cite{dean2010boron}.  This semi-classical (Drude) conductivity model is accurate as long as $\hbar \omega<<E_F$, and temporal modulation is  slow compared to any electron relaxation process. Graphene sheets with practical parameters usually fulfill these conditions in the THz regime.

\subsection{Surface currents on a periodic array of time-modulated graphene ribbons}

Let us first consider an infinite array of time-invariant (TI) graphene ribbons (with a constant Drude weight $W_D$) that is subject to a monochromatic TM plane wave. The structure has no variation in the $y$-direction, and due to the incidence of the TM-polarized wave, a surface current density will be induced on the ribbons in the $x$-direction. In the frequency domain, using phasor quantities, the total in-plane electric field phasor (at the arbitrary frequency $\omega$) is written as \cite{rahmanzadeh2021analytical}:
\begin{subequations}
\begin{equation}
\begin{split}
E_x(x)& = E_x^{ext}(x)+ \frac{1}{{j\omega {\varepsilon _0}}} \frac{d}{{dx}} \int\limits_{ - w/2}^{w/2} {G^0(x - x')} \frac{{d~{J_x}(x')~}}{{dx'}}dx'~ \\& - j\omega {\mu _0}\int\limits_{ - w/2}^{w/2} {G^0(x - x')} ~{J_x}(x')~dx'
\label{TI_IE}
\end{split}
\end{equation}
\begin{equation}
G^0(x - x') = \frac{1}{{4j}}\sum\limits_{l =  - \infty }^\infty  {H_0^{(2)}({k_0}\left| {x - x' - lD} \right|)} 
\label{TI_PGF}
\end{equation}
\label{TI_GR}
\end{subequations}
Here $E_x^{ext}(x)$ is the $x$-component of the incident electric field, $G^0(x - x')$ is the periodic free space Green’s function, and $k_0=\omega/c_0$ is the vacuum wave number with $c_0$ the speed of light. Note that the distribution of current on all ribbons is identical. Under the quasi-static approximation($k_0w<<1$ and $k_0D<<1$), \eqref{TI_GR} can be approximated by \cite{rahmanzadeh2019analytical}:
\begin{equation}
\begin{split}
    E_x(x)  = & E_x^{ext}(x) - \frac{1}{{2j\pi \omega {\varepsilon _0}}}\int\limits_{ - w/2}^{w/2} {\frac{{\partial G_1^0(x - x')}}{{\partial x}}} \frac{{d{J_x}(x')}}{{dx'}}dx'~ \\ & - j\omega {\mu _0}\int\limits_{ - w/2}^{w/2} {G_2^0(x - x')} ~{J_x}(x')~dx'
\end{split}
\label{TI_IE_qs}
\end{equation}
where
\begin{subequations}
\begin{equation}
\frac{{\partial G_1^{0}(x - x')}}{{\partial x}} = \sum\limits_{l =  - \infty }^\infty  {\frac{1}{{x - x' - lD}}} 
\label{TI_PGF_qs1}
\end{equation}
\begin{equation}
G_2^{0}(x - x')=\frac{c_0}{2 D} \frac{1}{j\omega}
\label{TI_PGF_qs2}
\end{equation}
\label{TI_PGF_qs}
\end{subequations}
%Here $c_0$ is the free space light speed. This equation was analytically solved in our previous papers using perturbation theory and eigenfunction expansion \cite{khavasi2014analytical}. 
Returning to the time domain, the total electric field on the graphene ribbons may be written as
\begin{equation}
\begin{split}
    E_x(x,t) = & E_x^{ext}(x,t) - \frac{1}{2\pi \varepsilon _0}\int\limits^{t}_{-\infty} \int\limits^{w/2}_{-w/2}  \frac{\partial  G_1^{0}(x - x')}{\partial x} \frac{\partial J_x(x',t')}{\partial x'} dx'dt \\ & - \frac{\eta _0}{2D}\int\limits^{w/2}_{-w/2} J_{x}(x',t)dx'
\end{split}
\label{TV_IE}
\end{equation}
where $\eta_0=\mu_0 c_0$ is the vacuum impedance. We next write the left hand side of the equation above (the total electric field) in terms of the surface current density using Eq.\  (\ref{Drude_TD}), and take the derivative of both sides of the resulting equation with respect to time. The resulting equation is

\begin{equation}
\begin{split}
 &\frac{\partial}{\partial t}   \left[ \frac{1}{W_D(t)}\left( \frac{\partial J_x(x,t)}{\partial t} +\gamma J_x(x,t) \right)\right]  \\=& \frac{\partial E_x^{ext}(x,t)}{\partial t}  - \frac{1}{{2\pi {\varepsilon _0}}}\sum\limits_{l =  - \infty }^\infty  {\int\limits^{w/2}_{-w/2} {\frac{1}{{x - x' - lD}}\frac{{\partial J(x',t)}}{{\partial x'}}dx'} } ~ \\&- \frac{\eta _0}{2D}\int\limits^{w/2}_{-w/2} \frac{\partial J_{x}(x',t)}{\partial t} dx' 
\end{split}
\label{TV_IE_qs}
\end{equation}
Take note that the above equation was obtained by applying the subwavelengtgh approximation to the frequency domain equation. Even though the incident wave is monochromatic, a time-varying system generates a spectrum of frequencies which differ from that of the incident wave. In order for the above equation to hold,  it is necessary that the subwavelength approximation remains valid at all generated new frequencies that are relevant to the problem ($\omega w/c_0 <<1$ and $\omega D/c_0<<1$). To solve \eqref{TV_IE_qs}, the induced surface current is expanded as
\begin{equation}
J_{x}(x,t) = \sum\limits_{n = 1} {{A_n}(t)} ~{\psi _n}(x)
\label{Jx_expan}
\end{equation}
where the functions $\psi_n(x)$ satisfy the eigenfunction equation 
\begin{equation}
\begin{split}
    \frac{1}{\pi }\sum\limits_{l =  - \infty}^\infty  {\int\limits_{ - w/2}^{w/2} {\frac{1}{{x - x' + lD}}\frac{{\partial {\psi _n}}}{{\partial x'}}~} } \partial x'=q_{n}\psi_{n}(x)
\end{split}
\label{EigenP_single}
\end{equation}
where $q_n$ are the corresponding eigenvalues. The functions $\psi_{n}$  satisfy the orthogonality condition 
\begin{equation}
\int\limits_{- w/2}^{w/2} \psi_{n}(x) \psi_{m}(x)=\delta_{m,n}
\end{equation}
where $\delta_{m,n}$ is the Kronecker delta. A method for calculating $\psi_{n}(x)$  and $q_n$ using Fourier expansion of  eigenfunctions was presented in \cite{khavasi2014analytical}. The first three eigenfunctions are listed in Table~\ref{eigenfunctions}. The higher-order eigenfunctions ($n > 3$) approximately equal $\sqrt{2/w} \cos(n\pi x/w)$ and $\sqrt{2/w} \sin(n\pi x/w)$ for odd and even orders, respectively. 
%Since the width of ribbons is small compared to all wavelengths involved ($k_0w<<1$), using perturbation theory, one can assume  to be the same eigenfunctions describing the induced surface currents on  time-invariant subwavelength graphene ribbons. These eigenfunctions satisfy \textcolor{red}{reference?}

\begin{table} 
\caption{The first three eigenfunctions for the problem of a time-invariant subwavelength graphene ribbon.}
\begin{center}
\begin{tabular}{c} 

\hline
 Eigenfunction \\ \hline
${\psi _1} = {w^{ - 0.5}}[1.2\sin \left( {\arccos {\kern 1pt} {{2x} \mathord{\left/
 {\vphantom {{2x} w}} \right.
 \kern-\nulldelimiterspace} w}} \right) - 1.06\sin \left( {3\arccos {{2x} \mathord{\left/
 {\vphantom {{2x} w}} \right.
 \kern-\nulldelimiterspace} w}} \right)] 
$ \\ \hline
 ${\psi _2} = {w^{ - 0.5}}[1.254\sin \left( {2\arccos {\kern 1pt} {{2x} \mathord{\left/
 {\vphantom {{2x} w}} \right.
 \kern-\nulldelimiterspace} w}} \right) - 0.302\sin \left( {4\arccos {{2x} \mathord{\left/
 {\vphantom {{2x} w}} \right.
 \kern-\nulldelimiterspace} w}} \right)]$  \\ \hline
\begin{array}{l}{\psi _3} = {w^{ - 0.5}}~[0.308\sin \left( {\arccos {\kern 1pt} {{2x} \mathord{\left/
 {\vphantom {{2x} w}} \right.
 \kern-\nulldelimiterspace} w}} \right) + 1.19\sin \left( {3\arccos {{2x} \mathord{\left/
 {\vphantom {{2x} w}} \right.
 \kern-\nulldelimiterspace} w}} \right)\\~~~~~~~~~~ - 0.484\sin \left( {5\arccos {{2x} \mathord{\left/ 
 {\vphantom {{2x} w}} \right.
 \kern-\nulldelimiterspace} w}} \right)]\end{array}

 \\ \hline
\end{tabular}
\end{center}
\label{eigenfunctions}
\end{table}

By substituting \eqref{Jx_expan} in \eqref{TV_IE_qs}, multiplying both sides of the equation by $\psi_n$, and carrying out an integration over the ribbon's width,  one arrives at
\begin{equation}
 \begin{split}
 &\frac{d}{d t}   \left[ \frac{1}{W_D(t)}\left( \frac{d A_n(t)}{d t} +\gamma A_n(t) \right)\right] +  \frac{q_n}{2\varepsilon _0} A_n(t)\\ 
 =& F_n(t)
 -\frac{\eta_0 S_n}{2D}\sum\limits_{m=1}^{\infty} S_m 
 \frac{d A_m(t)}{d t}
 \end{split}
\label{An}
\end{equation}
where
\begin{subequations}
\begin{equation}
S_n=\int\limits_{-w/2}^{w/2} \psi_n(x) dx
  \label{sn}
\end{equation}
\begin{equation}
F_n(t)=\int\limits_{-w/2}^{w/2} \frac{dE_x^{ext}(x,t)}{dt}\psi_n(x)
  \label{Fn}
\end{equation}
\end{subequations}
 The above system of coupled second-order ordinary differential  equations (ODE) for the coefficients $A_n(t)$ must, in principle, be solved together with initial conditions for $J_x$ (which can be translated into initial conditions for $A_n$). Note that the time-dependent coefficients $A_{n}(t)$ of the spatial eigenmodes $\psi_{n}(x)$ are coupled through the second term on the right hand side of \eqref{An}. This term emerges due to radiation of surface currents on graphene ribbons in the subwavelength ribbon array.
 
 In what follows, we assume that the external bias electric field applied to graphene ribbons is time-harmonic. This results  in  harmonic modulation of Drude weight as $W_D(t)=W_{D,0}[1+\alpha \cos\Omega t]$, where $\alpha$ and $\Omega$ are the strength and frequency of modulation, respectively. After switching on the dynamic modulation, once the stationary state is reached,  current density will contain various frequency components with the frequencies $\omega_k=\omega+k\Omega$ with $k$ an integer. The method of harmonic balance may then be used to find the steady state solution. To that end, the induced surface current and the incident monochromatic electric field are expressed as:
\begin{subequations}
\begin{equation}
{A_n}(t)~ = \frac{1}{2}\sum\limits_{k=-\infty}^{\infty} {{A_{n}^{k}}{e^{j{\omega _k}t}} + c.c} 
\label{An_hb}
\end{equation}
\begin{equation}
\label{Ex_hb}
E_x^{ext}(x,t) = \frac{1}{2}{E_0}~{e^{j{\omega _0}t}} + c.c
\end{equation}
\label{h_b}
\end{subequations}
where $A_{n}^{k}$ denotes the complex coefficient of $k'$th harmonic of $A_n(t)$ at $\omega_k$.  After substituting \eqref{h_b} in \eqref{An}, and equating the terms with equal frequencies on the two sides of  (\ref{An}), a system of linear equations is obtained for the coefficients $A_{n}^{k}$:
\begin{equation}
\begin{split}
& \sum_{l=-\infty}^{\infty}\left[ j\omega_{k}\left( j\omega_{l}+\gamma\right)\xi_{k-l} + 
\frac{q_n}{2\varepsilon _0}\delta_{k,l} \right]A_{n}^{l} \\ &=
j\omega_{0}E_{0} \delta_{k,0} -
j\omega_{k}\frac{\eta_0 S_n}{2D}\sum\limits_{m=1  }^{\infty} S_m A_{m}^{k}
\end{split}\label{eqan}
\end{equation}
where the coefficients $\xi_{k}$ are defined through the expansion
\begin{equation}
\frac{1}{W_D(t)}    =\frac{1}{W_{D,0}}\frac{1}{1+\alpha \cos\Omega t}=\sum_{k=-\infty}^{\infty} e^{jk\Omega t}\xi_k
\end{equation}
%As a further simplification, we next disregard the coupling between spatial modes of graphene ribbons by keeping the term with $m=n$ only in the summation on the right hand side of (\ref{equan}). This approximation is motivated by the observation that in the static case, i.e., in the absence of dynamic modulation,  \textcolor{red}{reference?} 

Since $\alpha$ is always smaller than unity ($|\alpha|<1$), $\xi_{k}$ can approximately be calculated using the Taylor series. For small values of $\alpha$, approximated expressions for $\xi_{k}$ are presented in Table~\ref{Xi_k}.

\begin{table} 
\caption{Approximated expressions for $\xi_{k}$.}
\begin{center}
\begin{tabular}{|c|c|} 

\hline $\xi_{0}\approx\frac{1}{W_{D,0}}(1+\frac{1}{2}\alpha^2+\frac{3}{8}\alpha^4)$ & $\xi_{\pm3}\approx\frac{-\alpha^3}{8W_{D,0}}$ \\ \hline  $\xi_{\pm1}\approx\frac{-1}{2W_{D,0}} (\alpha+\frac{3}{4}\alpha^3)$ & $\xi_{\pm4}\approx\frac{\alpha^4}{16W_{D,0}}$
\\ \hline $\xi_{\pm2}\approx\frac{1}{4W_{D,0}}(\alpha^2+\alpha^4)$ &
$\xi_{\pm k} \approx 0$ ~~ for~~ $|k|>4$  \\  \hline
\end{tabular}
\end{center}
\label{Xi_k}
\end{table}

It is instructive to inspect the solution of (\ref{eqan}) when graphene ribbons are not dynamically modulated. In that case $\xi_k=\xi_0 \delta_{k,0}$ with $\xi_0 =1/W_{D,0}$. All coefficients are zero except $A_{n}^{0}$ which satisfy
\begin{equation}
A_{n}^{0}=\frac{j\omega_{0}W_{D,0}}{\nu_{n}^{2}-\omega_{0}^{2} +j\omega_{0}\gamma} \left( E_{0} -
\frac{\eta_0 S_n}{2D}\sum\limits_{m=1}^{\infty} S_m A_{m}^{0}\right)\label{equan2}
\end{equation}
where 
\begin{equation}
    \nu_{n}^{2}=\frac{q_n W_{D,0} }{2\varepsilon _0}
\end{equation}
Solution of these equations show that if the frequency $\omega_0$ of the incident plane wave is relatively close to $\nu_p$ for for a particular (spatial) mode $p$, resonance behavior is observed and the corresponding amplitude $A_{p}^{0}$ becomes much larger than that of other modes $A_{n}^{0}, n\neq p$. Moreover, neglect of the terms with $m\neq n$ in the summation on the right hand side of (\ref{equan2}) will not significantly affect the results. 

In presence of harmonic modulation, an incident wave will produce a discrete spectrum of frequencies $\omega_k =\omega_0 + k \Omega$. If the modulation is not too strong, one expects the amplitude of these additional frequencies to drop fast with increasing $k$. If, in addition, $\Omega << \omega_0$, the produced frequencies will be effectively close to $\omega_0$. Therefore, if $\omega_0$ is again relatively close to some $\nu_p$, we expect the corresponding amplitudes $A_{p}^{k}$ to be dominant and also may disregard the coupling terms. 

The calculation presented can be generalized to the case where ribbons are placed on an arbitrary layered substrate. In Appendix A, we present the generalization of the proposed method for the case that ribbons are sandwiched between two homogeneous half spaces with relative permittivities $\varepsilon_{r1}$ and $\varepsilon_{r2}$ (Fig~\ref{GR_TwoHalf}).  Furthermore, similar to \cite{rahmanzadeh2019analytical}, generalization to oblique incidence is possible with the correction of Green's function. However, we do not present details of calculations for oblique incidence for brevity.

\subsection{Reflection and transmission coefficients}
\begin{figure}
\centering \includegraphics[width=\linewidth]{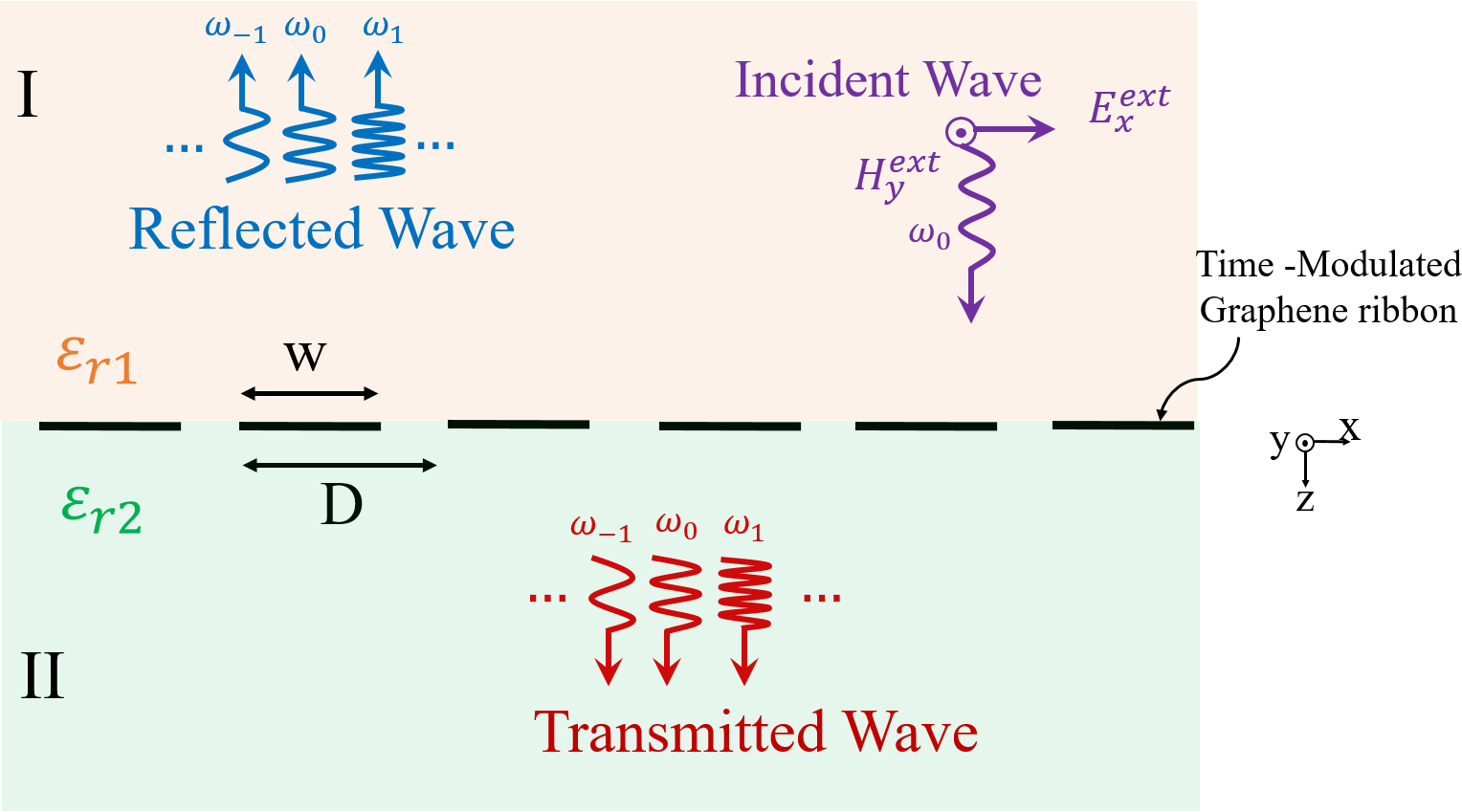}
\caption{Graphene ribbons are sandwiched between two homogeneous isotropic media with relative permittivities $\varepsilon_{r1}$ and $\varepsilon_{r2}$. A monochromatic transverse plane wave normally impinges on time-modulated graphene ribbons from medium I toward medium II.}
\label{GR_TwoHalf}
\end{figure}

In this subsection, we use the derived current distribution to calculate reflection and transmission coefficients. To that end, we apply the Rayleigh expansion, generalized to include various harmonics, to the total field above and below the ribbon array. The electromagnetic field in the structure shown in Fig~\ref{GR_TwoHalf}, is expressed as 
\begin{subequations}
\begin{equation}
   {H_{1y}} =\frac{1}{2}[ {e^{ - jk_{z,0}^{(1)}z + j{\omega_0}t}} +\sum\limits_p {{R_p}{e^{jk_{z,p}^{(1)}z + j{\omega _p}t}}}] + c.c
   \label{Hy1}
\end{equation}
\begin{equation}
{E_{1x}} = \frac{1}{2}[\xi _0^{(1)}{e^{ - jk_{z,0}^{(1)}z + j{\omega_0}t}} -\sum\limits_p {\xi _p^{(1)}{R_p}{e^{jk_{z,p}^{(1)}z + j{\omega_p}t}}}]+ c.c  
  \label{EX1}
\end{equation}
\label{E_H1}
\end{subequations}
for the region I ($z<0$), and
\begin{subequations}
\begin{equation}
   {H_{2y}} = \frac{1}{2}~\sum\limits_p {{T_p}{e^{ - jk_{z,p}^{(2)}z + j{\omega _p}t}}}+ c.c
   \label{Hy2}
\end{equation}
\begin{equation}
 {E_{2x}} = \frac{1}{2}\sum\limits_p {\xi _p^{~(2)}~{T_p}{e^{ - jk_{z,p}^{(2)}z + j{\omega _p}t}}}  + c.c  
  \label{Ex2}
\end{equation}
\label{E_H2}
\end{subequations}
for the region II ($z>0$). Here $R_p$ and $T_p$ are the reflection and transmission coefficients at $\omega_p$, respectively, and
\begin{subequations}
\begin{equation}
   k_{z,p}^{(j)} = \sqrt {{\varepsilon _{rj}}}~\omega _p/c_0~~;~~{\omega _p} = {\omega _0} + p\Omega
   \label{kzp}
\end{equation}
\begin{equation}
\xi _p^{(j)} = \frac{{k_{z,p}^{(j)}}}{{{\omega _p}{\varepsilon _0}{\varepsilon _{rj}}}} 
  \label{Xip}
\end{equation}
\label{k-Xi}
\end{subequations}

Next, we apply the electromagnetic boundary conditions at $z=0$,
\begin{subequations}
 \begin{equation}
  E_{x1}=E_{x2}    
 \end{equation}
    \begin{equation}
    H_{1y}-H_{2y}=J_x
    \end{equation}
    \label{BC}
\end{subequations}
After substituting \eqref{E_H1} and \eqref{E_H2} in \eqref{BC}, the reflection and transmission coefficients are obtained after some mathematical manipulations as
\begin{subequations}
\begin{equation}
{T_0} = \frac{{\xi _0^{~(1)}}}{{\xi _0^{~(2)}}}(1 - {R_0}) = \frac{{{{2\xi _0^{~(1)} - \xi _0^{~(1)}}/D}}}{{\xi _0^{~(1)} + \xi _0^{~(2)}}}\sum\limits_{n = 1} {{A_{~0,n}}} ~{f_n}~\
   \label{T0}
\end{equation}
\begin{equation}
 T_p =  - \frac{{\xi _p^{~(1)}}}{{\xi _p^{~(2)}}}{R_p} =  - \frac{{\xi _p^{~(1)}}}{{D(\xi _p^{~(1)} + \xi _p^{~(2)})}}\sum\limits_{n = 1} {{A_{p,n}}} ~{f_n}~;~~p \ne 0 
  \label{Tp}
\end{equation}
\label{T}
\end{subequations}

Please note that, in this paper, we assumed a monochromatic excitation. However, response to an arbitrary excitation could be found using the Fourier series expansion of the incident field since the structure is linear, as long as the subwavelength approximation holds.

\section{Numerical results and discussion}

\begin{figure*}
\centering \includegraphics[width=\linewidth]{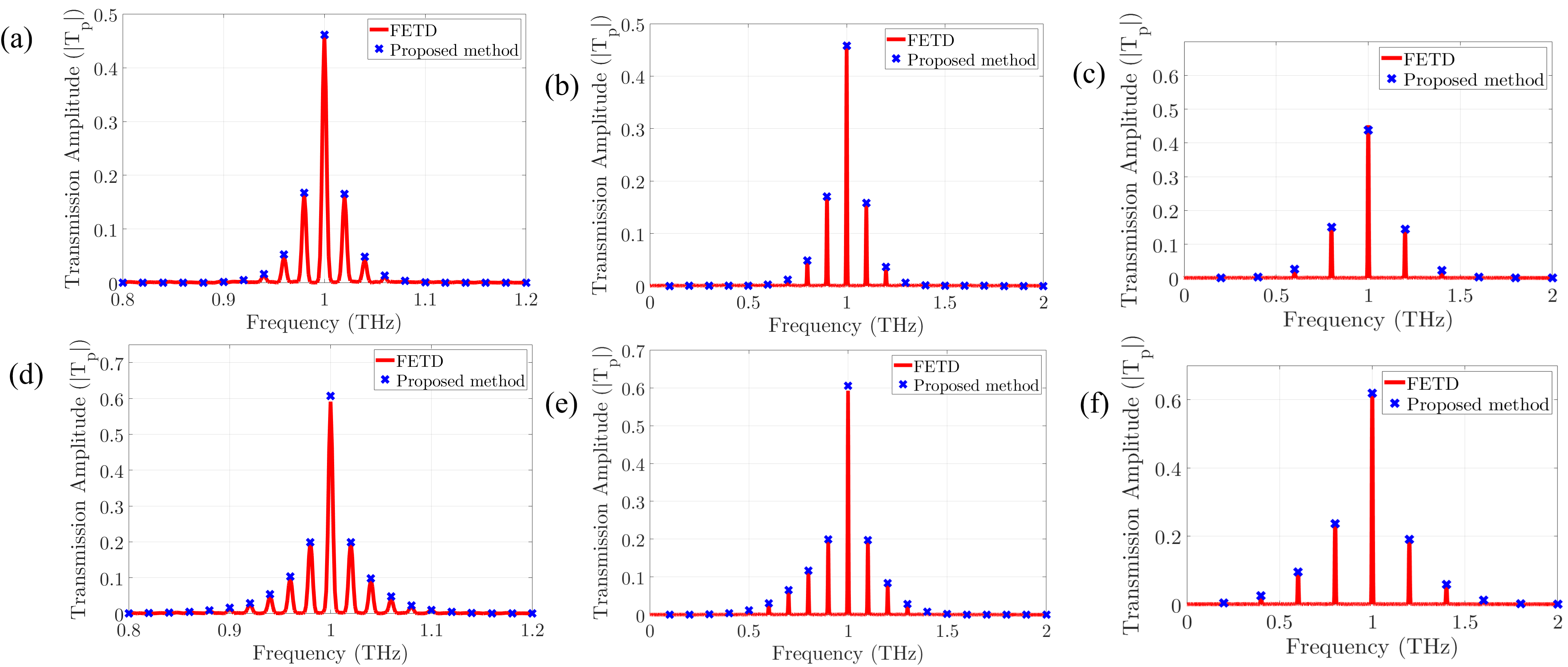}
\caption{Comparison of the results of the proposed method with full-wave simulations (FETD). The amplitudes of transmitted waves are plotted for different modulation parameters. (a)-(c) $\alpha=0.3$ and $\Omega$ is 20GHz, 100GHz, and 200GHz, respectively.  (d)-(e) $\alpha=0.6$ and $\Omega$ is 20GHz, 100GHz, and 200GHz, respectively. The other structure parameters are $\omega=2\pi\times 1$THz, $D=60\mu$m, $w=42\mu$m, $E_{F0}=0.135$eV, $\tau=1p$s, and $\varepsilon_{r1}=$$\varepsilon_{r2}=1$.}
\label{ex1}
\end{figure*}

\begin{figure}
\centering \includegraphics[width=0.75\linewidth]{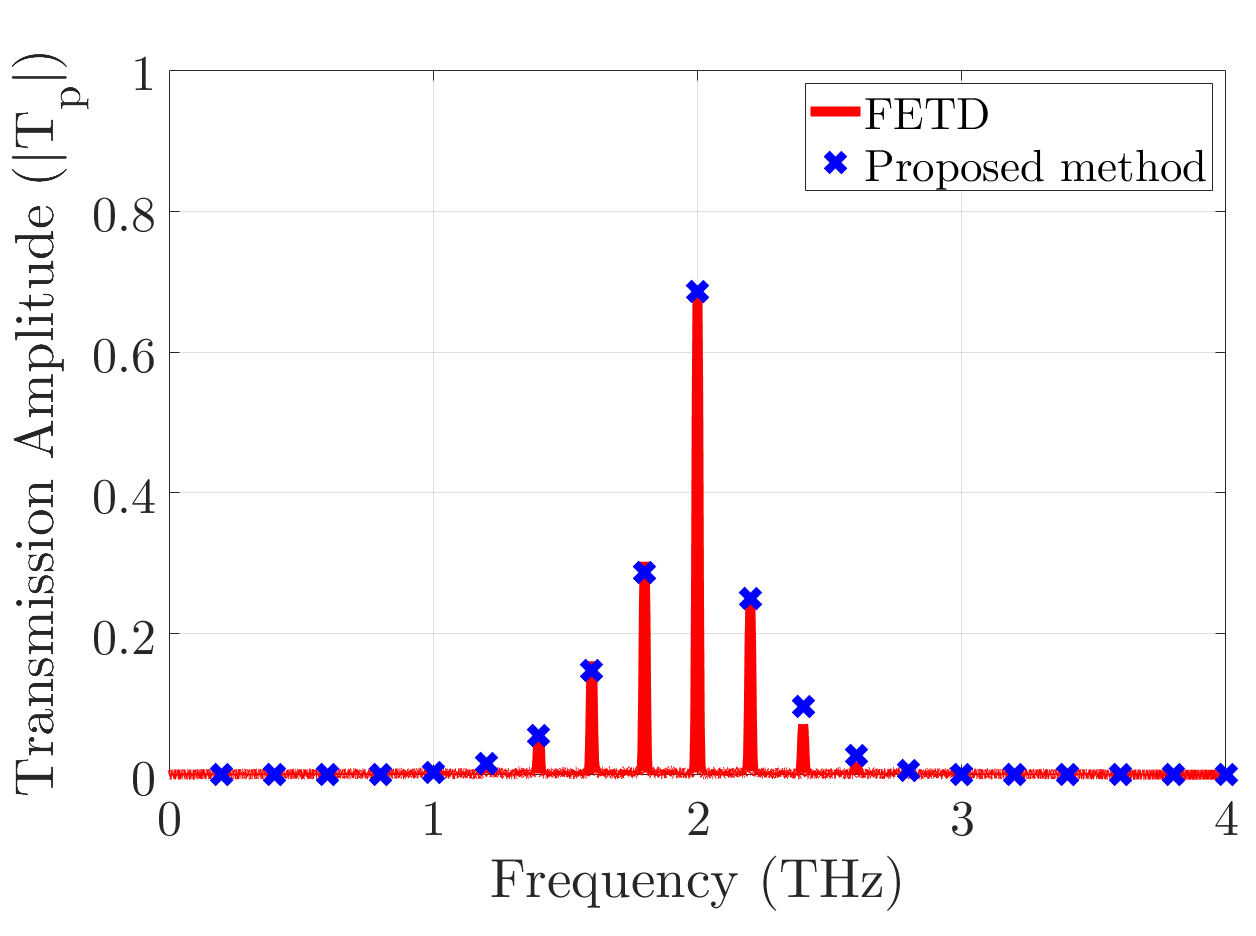}
\caption{Frequency-domain response of the structure for a monochromatic TM-plane wave ($\omega=2\pi\times 2$THz). The time-modulated graphene ribbon parameters are  $D=12\mu$m, $w=9\mu$m, $E_{F0}=0.2$eV, $\tau=1p$s, $\alpha=0.4$, $\Omega=2\pi\times 200$GHz, $\varepsilon_{r1}=1$, and $\varepsilon_{r2}=2.25$.}
\label{ex2}
\end{figure}

In this section we verify the accuracy of our method through some numerical examples. As the first example, consider a periodic array of time-modulated graphene ribbons with $D=60\mu$m, $w=42\mu$m, $\varepsilon_{r1}=\varepsilon_{r2}=1$, $\tau=1p$s, and $E_{F0}=0.135$eV. Using \eqref{T} the absolute value of transmission coefficients is calculated and plotted in Fig.~\ref{ex1} for different strengths and frequencies of modulation.  As the next example, we consider a periodic array of time-modulated graphene ribbons mounted on a dielectric substrate with the permittivity $\varepsilon=2.25\varepsilon_0$ (substrate thickness is assumed to be infinite for simplicity). Graphene parameters and geometric parameters are  $D=12\mu$m, $w=9\mu$m, $\alpha=0.4$, $\Omega=2\pi\times$200GHz, $E_{F0}=0.2$eV, and $\tau=1p$s. The magnitude of transmission coefficients is plotted in Fig.~\ref{ex2}. The structure is illuminated by a monochromatic TM-plane wave with a frequency of 2 THz. Figs.~\ref{ex1} and \ref{ex2} also show the corresponding finite-element-time-domain (FETD) results. Excellent agreement is observed between the two approaches. The FETD results were obtained using the commercial program COMSOL Multiphysics 5.6. (Appendix B presents a detailed discussion of carrying out FETD simulations by COMSOL.) The simulations were performed on a computer with an Intel(R) Core(TM) i7-6700HQ CPU and an installed memory (RAM) of 16.00 GB. In this case, it took only 13 ms and 14 ms on average to calculate the transmission coefficients using the proposed method for the first and second examples, respectively. By contrast, FETD simulations took about an hour on the same computer. It should be noted that in the subwavelength regime it is usually sufficient to include the first and third eigenfunctions in the calculation as effects of higher order eigenfunctions are negligible (at normal incident the even modes are absent because $S_n$ vanishes in this case). The results presented in Figs.~{\ref{ex1}} and {\ref{ex2}} were obtained considering just the first eigenfunction since the operation frequency is close  to the first resonance frequency of graphene ribbons (The first resonance frequency occurs at 1 THz and 2.1 THz for the first and second examples, respectively).

The monochromatic incident wave is reflected as plane waves with different frequencies as shown in Figs.~\ref{ex1} and \ref{ex2}. A comb-like shape in transmission coefficients emerges. In \cite{menendez2017frequency} a similar frequency comb was reported which was analyzed  using coupled mode theory (CMT). However, our proposed method has several advantages and features that CMT lacks. \cite{menendez2017frequency} assumes that graphene conductivity is dispersionless, which is correct in a limited frequency band. By contrast, our proposed method can predict the response of the structure in a wide range of frequencies. Besides, much higher modulation frequencies are allowed by our method and unlike CMT, there are no  unknown coefficients to be found by fitting to numerical results. Finally, our proposed method can identify the physical principles behind the structure behavior, which is not available in the CMT. Some of these physical principles have been discussed in subsection II.b. Furthermore, from Figs.~{\ref{ex1}} and {\ref{ex2}}, it can be seen that when the harmonic number ($k$) increases, the energy transferred to the harmonic decreases. This point can be explained from \eqref{eqan} and Table~\ref{Xi_k}. Since $\alpha$ is smaller than unity, $\alpha^{|k|} > \alpha^{|k|+a}$ ($a$ is a positive integer). Since $\xi_k$ is predominantly determined by the term which is proportional to $\alpha^{|k|}$, magnitude of $\xi_{\pm k\mp a}$ is smaller than that of $\xi_{\pm k}$.   Therefore, from \eqref{eqan} it may be concluded that an increase in the harmonic number leads to a decrease in the magnitude of $A_n^{k}$ and leads to a smaller portion of the energy transferred to the higher-order harmonic.

Note that the parameters of graphene ribbons were arbitrarily chosen and are not optimized. Using our proposed method and optimizing the parameters, one can design a frequency comb generator and other devices with better features than previously reported devices. Nevertheless, designing novel devices falls out of the scope of this paper.

\begin{figure}
\centering \includegraphics[width=\linewidth]{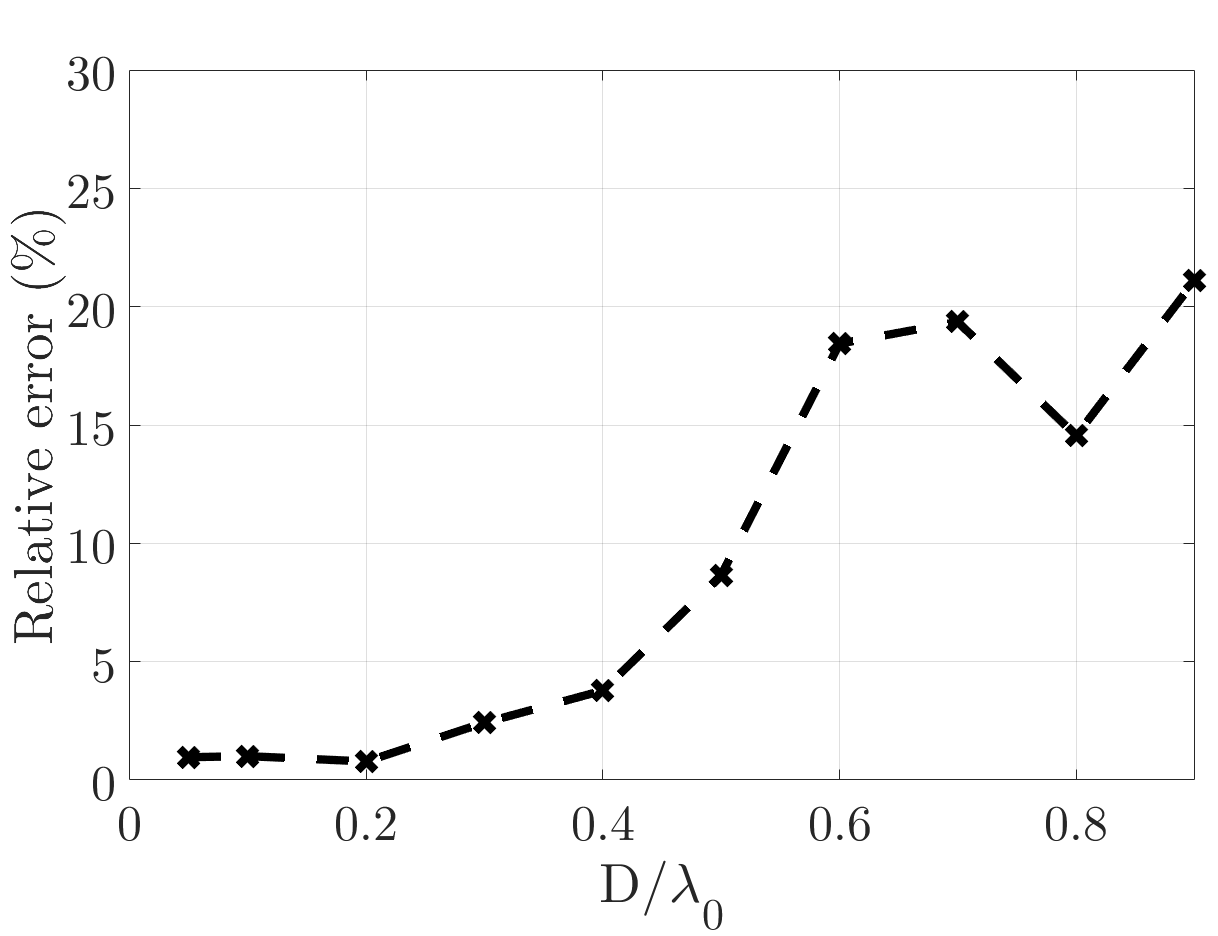}
\caption{Relative errors in calculating the zero-order transmission coefficient for different values of $D/\lambda$. The frequency of excitation is 1 THz, and the array constant is assumed to change between 15 and 270$\mu$m. Other structure parameters are  $w=0.7D$, $\alpha=0.4$, $\Omega=\omega/10$, $\tau=1p$s, and $\varepsilon_{r1}=$$\varepsilon_{r2}=1$.}
\label{D-lambda}
\end{figure}

\begin{figure}
\centering \includegraphics[width=\linewidth]{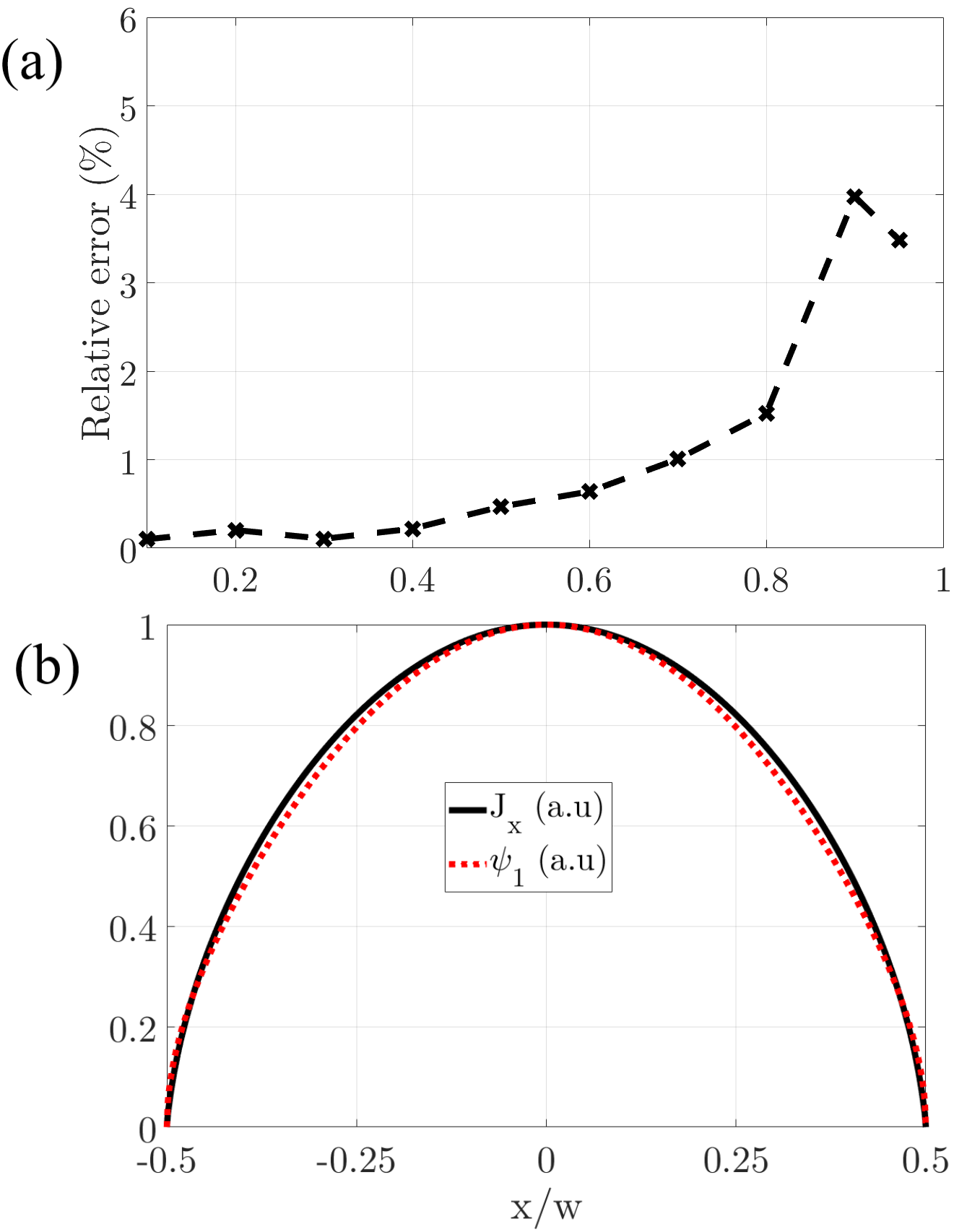}
\caption{(a) Relative errors in calculating the zero-order transmission coefficient for different values of $w/D$. The time-modulated graphene ribbons parameters are  $\omega =2\pi\times 1$THz, $D=30\mu$m, $\alpha=0.4$, $\Omega=0.1 \omega$,  $\tau=1p$s, and $\varepsilon_{r1}=$$\varepsilon_{r2}=1$. (b) Surface current distribution at the vicinity of the first resonance frequency for the same structure with $w/D=0.7$ and $\psi_1$.}
\label{w/D}
\end{figure}

Finally, let us discuss the limitations of the proposed method. We used quasi-static approximation to arrive at equation \eqref{TV_IE_qs}. Consequently, the operating wavelength must be larger than the width and period of the graphene ribbon array (subwavelength limit). Therefore, we expect that when $w/\lambda$ or $D/\lambda$ increases, the the proposed method becomes less accurate. To better appreciate the role of this approximation, the relative error in the calculation of the zero-order transmission coefficient of the structure is plotted as a function of $D/\lambda$ in Fig.~\ref{D-lambda}. Fig.~\ref{D-lambda} shows that the proposed method has acceptable accuracy for $D/\lambda < 0.4$, which demonstrates its capability for designing different applications since graphene ribbons are usually utilized in this range. Note also that in \eqref{Jx_expan}, an infinite number of eigenfunctions is, in principle, needed.  These eigenfunctions, which must satisfy \eqref{EigenP_single}, are approximated by those of a static single ribbon\cite{khavasi2014analytical}. Increasing the fill factor ($w/D$) leads to stronger interaction between neighboring ribbons and reduces the accuracy of eigenfunctions (contrary to eigenvalues which are corrected). The expansion \eqref{Jx_expan} does not depend on the nature of the functions $\psi_n$ as long as they constitute a complete set.  In practice, however, only a few terms are used and if these few terms do not reproduce the current distribution in the actual problem error increases.  For  better assessment  of this effect, the relative error in  the zero-order transmission coefficient of a periodic array of graphene ribbons is plotted as a function of the fill factor in Fig.~\ref{w/D}(a). It is observed that by increasing the fill factor, the relative error in the prediction of the magnitude of transmission will slightly increase. We also plot the current distribution on the time-modulated graphene ribbon near its first resonance in Fig.~\ref{w/D}(b). The current distribution is obtained by FETD simulation and is compared with corresponding eigenfunctions (n=1). Good agreement is observed between the results, which implies that the perturbation approximation has no significant effect on the accuracy of our proposed method. Besides, the time-domain Drude model has some limitations on graphene parameters which are noted in Section II. However, these limitations are not related to our proposed method, and with correction in the time domain model of graphene (correction on the left-hand side of \eqref{TV_IE_qs}), our method may be applicable in these cases.

\section{Conclusion}
Based on time-domain integral equations governing the surface current density on time-modulated graphene ribbons, a semi-analytical, fast, and accurate method is presented in this paper. Unlike previous studies, the proposed method does not suffer from the extremely time-consuming nature of common numerical techniques, allows much higher modilations frequencies, and can be useful for understanding the electromagnetic behavior of the structure. We first derived a semi-analytical expression for the surface current density of graphene ribbons illuminated by a TM polarized plane wave under subwavelength approximation. The reflection/transmission coefficients were obtained using the electromagnetic field expansion and appropriate boundary conditions. Various numerical examples demonstrated the accuracy and applicability of the proposed method. FETD results showed that our method has only a subwavelength limitation (the periodicity of the structure and ribbon's width must be approximately less than $0.3\times$the wavelength), which is not a significant problem because graphene ribbons are usually used in this regime. Our work presents a new method for analyzing time-modulated graphene ribbons, which is more precise, rapid, and affordable than the methods in the literature. Our results also offer a deeper insight into the physical properties of time-modulated graphene ribbons, which may be useful for devising different novel devices.

\appendices
\section{Effect of dielectric half-space}

By modifying Green's function and the external field, the proposed method for calculating induced surface current can be generalized to those cases where graphene ribbons are placed in a multi-layered structure. Here we  present calculations for the case where the graphene ribbons are sandwiched between two half-spaces. The upper medium ($z<0$) and the lower medium ($z>0$) are filled with a homogeneous isotropic dielectric with relative permittivities $\varepsilon_{r1}$ and $\varepsilon_{r2}$, respectively as displayed in Fig.{\ref{GR_TwoHalf}}.
We apply three modifications to \eqref{TV_IE_qs} to derive the induced surface current. The first term on the right-hand side of \eqref{TV_IE_qs} must be replaced with the incident field plus the field reflected by the lower half-space in the absence of graphene ribbons. The second term stemmed from the scalar potential and has an electrostatic nature; hence, it is sufficient  to replace $\varepsilon_0$ with the effective permittivity ($\varepsilon_{eff}=\varepsilon_0(\varepsilon_{r1}+\varepsilon_{r2})/2$). The last term is modified as:
\begin{equation}
-\frac{\eta_0}{D( \sqrt{\varepsilon_{r1}}+\sqrt{\varepsilon_{r2}})}\int\limits^{w/2}_{-w/2} \frac{\partial J(x',t)}{\partial t} dx' 
\label{GP_TD_sub}
\end{equation}
which is obtained by applying the inverse Fourier transform to the zero-order term in the expansion of periodic Green's function (corresponding to the subwavelength array) presented in \cite{rahmanzadeh2021analytical}. Like Section II, $J_x(x,t)$ can be expanded in terms of the corresponding eigenfunctions and unknown complex coefficients and a  system of equations similar to \eqref{eqan} is found, by solving which the surface current can be calculated.

\section{Time-modulated graphene simulations in COMSOL}
FETD results are carried out using the transient electromagnetic wave (temw) module combined with a coefficient from the boundary PDE (CB) module in COMSOL 5.6. We added a surface current in the temw module for modeling graphene ribbons. We also set the time-domain Drude equation in the CB module. These modules are linked through the surface current boundary condition, so we can simultaneously solve Maxwell's equation and the time-domain Drude equation. A simulation box of $D \times 1\lambda_0$ (where $\lambda_0$ is the free space wavelength of the excitation signal) was implemented, and periodic boundary (continuous) conditions were applied in the $x$-direction. Moreover, a scattering boundary condition was used at the top and bottom of the simulation box to avoid undesirable reflection. The incident filed ($E_x^{ext}$) is a sinusoidal wave with a wide Gaussian envelope in the time domain (monochromatic Gaussian pulse) applied from the top boundary of the simulation box via the scattering boundary condition. The duration input pulse is assumed to be $700 T_0$ ($T_0=1/f_0$ and $f_0$ is the frequency of excitation), and its expression can be written as 
\begin{equation}
    E_x^{ext}(t)=E_0 \cos{(\omega_0 t)} \exp{[-(t-t_0)^2/{\Delta t}^2}]
    \label{Guas}
\end{equation}
where $\omega_0=2\pi f_0$ is the pulse center frequency, $\Delta t=700/\omega_0$ gives the width of the pulse, and $t_0=3 \Delta t$ is the pulse center time. A triangular mesh was implemented with a maximum size of $D/6$. To better ensure the accuracy of FETD, we set the method of time stepping to generalized alpha in COMSOL. Finally, we applied the fast Fourier transform (FFT) to the time domain result and obtained the transmission coefficient by calculating the ratio of the field at the bottom of the simulation box to the incident field in the frequency domain.

\section*{Acknowledgment}
We would like to thank Dr. Emanuele Galiffi for helping us perform the FETD simulations in COMSOL. 

\bibliographystyle{IEEEtran}
\bibliography{sample}
\end{document}